# Enhancing Data-driven Multiscale Topology Optimization with Generalized De-homogenization

Liwei Wang [a,b], Zhao Liu [c], Daicong Da [b], Yu-Chin Chan [b], Wei Chen [b], Ping Zhu [a]

a. The State Key Laboratory of Mechanical System and Vibration, School of Mechanical Engineering, Shanghai Jiao Tong University
Shanghai, P.R. China

b. Dept. of Mechanical Engineering, Northwestern University
Evanston, IL, USA

c. School of Design, Shanghai Jiao Tong University
Shanghai, P.R. China

**Abstract**

De-homogenization is becoming an effective method to significantly expedite the design of high-resolution multiscale structures, but existing methods have thus far been confined to simple static compliance minimization problems. There are two critical challenges to be addressed in accommodating general design cases: enabling the design of unit-cell orientation and using free-form microstructures. In this paper, we propose a data-driven de-homogenization method that allows effective design of the unit-cell orientation angles and conformal mapping of spatially varying, complex microstructures. Instead of using conventional square cells with rectangular holes, we devise a parameterized microstructure composed of rods in different directions to provide more diversity in stiffness while retaining geometrical simplicity. The microstructural geometry-property relationship is then surrogated by a multi-layer neural network to avoid homogenization analysis during optimization. A Cartesian representation of the unit-cell orientation is incorporated into homogenization-based optimization to design the angles. Corresponding high-resolution multiscale structures are obtained from the homogenization-based designs through a conformal mapping constructed with sawtooth function fields. This allows us to assemble complex microstructures with an oriented and compatible tiling pattern, while preserving the local homogenized properties. To demonstrate our method with a specific application, we optimize the frequency response of structures under harmonic excitations within a given frequency range. It is the first time that a sawtooth function is applied in a de-homogenization framework for complex design scenarios beyond static compliance minimization. The examples illustrate that high-resolution multiscale structures can be generated with high efficiency and much better dynamic performance compared with the macroscale-only optimization. Beyond frequency response design, our proposed framework can be applied to general static and dynamic problems.

# 1. Introduction

Recent advances in computational and manufacturing capability have fueled the development of topology optimization methods for multiscale structures [1]. In multiscale design, the structure is optimized at both macro- and micro-scales to provide unprecedented flexibility in meeting spatially varying properties, leading to the desired structural performance. However, multiscale microstructure design suffers from computational challenges that deprive it of practical applications, including exhaustive computational cost, high-dimensional design variables, and unconnected microstructures. These issues become even more critical for design cases involving dynamic, nonlinear, and multi-physics behaviors. The aim of this study is thus to develop a new data-driven approach that can mitigate these computational challenges to expedite multiscale topology optimization without much sacrifice to its design flexibility.

Existing multiscale topology optimization methods can be largely classified into four categories based on the tiling patterns they assume for the microstructures, i.e., periodic, aperiodic, multi-domain, and functionally graded designs. Periodic designs assume the full structure is assembled by the same microstructure, enabling a highly efficient design process at the cost of suboptimal designs [2-6]. In contrast, aperiodic designs associate each region in the discretized full structure with an independent microscale optimization, which can significantly increase the design freedom to obtain better design performance [7, 8]. However, the large number of microstructures to be designed, expensive multiscale performance calculations, and a nested optimization scheme lead to an unaffordable computational cost. Also, it is difficult to ensure compatible neighboring microstructures in the aperiodic tiling, which is critical for structural integrity and manufacturability [9, 10]. Multi-domain designs divide the full structure into several subregions of periodic microstructures, compromising between the efficiency of periodic designs and the design flexibility of aperiodic designs [11-15]. The compatibility between microstructures is considered in the optimization by prespecifying connectors or adding extra constraints [12, 16]. Nevertheless, multi-domain designs are often confined to a small number of subregions in order to keep computational costs within an acceptable range. As an alternative to reach a trade-off between periodic and aperiodic designs, functionally graded designs assume microstructures within the full structure share the same architecture but with varying geometrical parameters [17-20]. A subsequential advantage brought by this assumption is that one could leave out the microscale details during the optimization and directly optimize the spatial distribution of geometrical parameters at the macroscale. After optimization, the corresponding multiscale structure can be obtained by filling elements in the macroscale design with microstructures specified by the optimized parameters. The former optimization process is called homogenization-based design while the latter microstructure tiling process is known as de-homogenization. Moreover, this type of design is amenable to the data-driven framework, which exploits an efficient surrogate model to replace the expensive on-the-fly homogenization. The functionally graded tiling could also ensure smooth transitions between neighboring microstructures for better compatibility. To further increase the design freedom, multiple microstructure architectures can be considered by integrating discrete material optimization [21-23] or latent-variable mapping [24-26]. Despite the promising results, most existing functionally graded designs fail to take into account the change of microstructure orientation, which significantly restricts the range of achievable properties and therefore leads to suboptimal performance.

The major obstacle in designing graded structures with oriented microstructures lies in the de-homogenization process. When the unit-cell orientations vary across the macroscale structure, the corresponding microstructures need to be rotated accordingly to assemble the full structure. As a result, neighboring microstructures might not be connected with each other after rotation, making the multiscale structure fail to attain the designed performance and even become unmanufacturable. In a major advance in 2008 [27], Pantz and Trabelsi focused on square microstructures with rectangular holes and proposed a method to project a homogenized design to a multiscale oriented structure on a high-resolution mesh. They achieved this de-homogenization process via implicit mapping functions constructed through least-squares minimization. Later, Groen and Sigmund [28] simplified this de-homogenization process by introducing the connected component labeling method to obtain a consistent orientation field, and relaxing the optimization problem for the mapping function. This simplified method was further extended to enable the efficient design of 3D multiscale structures [29, 30]. Meanwhile, several modifications have been proposed to address singularity issues in orientation field design that would otherwise cause de-homogenization to fail [31, 32]. Although the de-homogenization method is appealing, it is still confined to simple static compliance minimization problems [33, 34]. The reason is that it can only handle square cells with rectangular holes, simply making the unit-cell orientation align with the principal strain direction. While these designs are optimal for compliance minimization given a single loading, they would become suboptimal for general design cases, such as multi-loading, dynamic response optimization, and multi-physics problems.

In the last few years, attempts have been made to incorporate orientation design (unit-cell orientation as optimization variables) and free-form microstructure mapping into de-homogenization. Kim and Lee et al. [35, 36] developed an explicit de-homogenization method that allows the design of unit-cell orientation, instead of simply aligning it with the principle strain direction. Similar to the texture mapping process, Groen [37] directly used cosine fields to construct the mapping functions in de-homogenization for complex microstructures that could be described as a parallelogram or parallelepiped. However, this mapping is not conformal (angle-preserving), which will lead to the deviation of structural performance from its homogenization-based design. Geoffroy [38] devised stochastic microstructures with a special parameterization scheme to allow the direct use of de-homogenization, which is not applicable for general microstructure designs. Meanwhile, Tamijani et al. [39] decomposed the complex unit-cell microstructure into Fourier series and then constructed a mapping function for each of the spatial harmonics. While it allows the mapping of free-form microstructures, it needs to calculate a large number of mapping functions (up to 127 in [39]), which is complex and time-consuming. Kumar et al. [40] constructed a finite element model to solve the oriented mapping problem for complex microstructures, which overly complicates the implementation process. As a result, no one, to the best of our knowledge, has successfully integrated orientation design and free-formed microstructure mappings with de-homogenization to enable efficient multiscale design for applications beyond simple static compliance minimization. Developing a method that can accomplish this is the goal of this study.

Specifically, as shown in Fig. 1, our approach first devises a parameterized unit-cell composed of bar groups aligned in different directions to achieve a wide range of properties. This parametrized unit-cell is then used to construct a large database with precomputed homogenized properties, i.e., stiffness tensor and volume fraction, from which a neural network can be trained to surrogate the geometry-property relations. This surrogate model is integrated into a homogenization-based optimization, with both geometrical parameters and unit-cell orientation as design variables. We then propose to use sawtooth function fields [41] to construct a conformal mapping from free-form rectangular unit-cells to a microstructure tiling with the prescribed orientation, i.e., an ensemble of microstructures covering a given region. This sawtooth-function-based mapping ensures good connections between oriented microstructures and can preserve the original homogenized properties, due to its unique conformality. It is used to extend the de-homogenization process in accommodating spatially varying, complex unit-cell microstructures to obtain multiscale structures corresponding to the homogenization-based designs.

While the proposed framework is applicable to various design applications, we found it particularly useful for frequency response optimization, which can improve both the efficiency and performance of the design process. Minimizing the frequency response of a structure under a given excitation is important for engineering design to suppress undesirable vibrations [42-44]. Existing TO methods for frequency response optimization mainly focus on macroscale designs and efficient reduced-order models to accelerate the frequency response calculation [43, 45-51]. Due to the exhaustive computational cost, few studies have realized multiscale TO for frequency response optimization, and the solutions are confined to periodic [4, 52, 53] and functionally graded designs with fixed unit-cell orientation [19]. Moreover, unlike static compliance minimization, the structure in frequency response optimization will undergo different distortion states within the excitation frequency range of interest. Therefore, square unit-cells with rectangular holes aligned with the principal strain direction, which are adopted in most de-homogenization methods, might not be optimal or even applicable for frequency response optimization. In this study, through multiple design cases, we demonstrate that the proposed data-driven method incorporated with sawtooth-function-based de-homogenization can provide superior efficiency and flexibility that can benefit frequency response optimization. It successfully generates high-resolution multiscale structures with efficiency comparable to single-scale macroscale design while achieving better dynamic performance. To the authors' knowledge, this is the first time that the de-homogenization method has been applied to design applications other than static compliance optimization. With the enhanced flexibility, the proposed method can also accommodate other general static and dynamic problems beyond frequency response optimization.

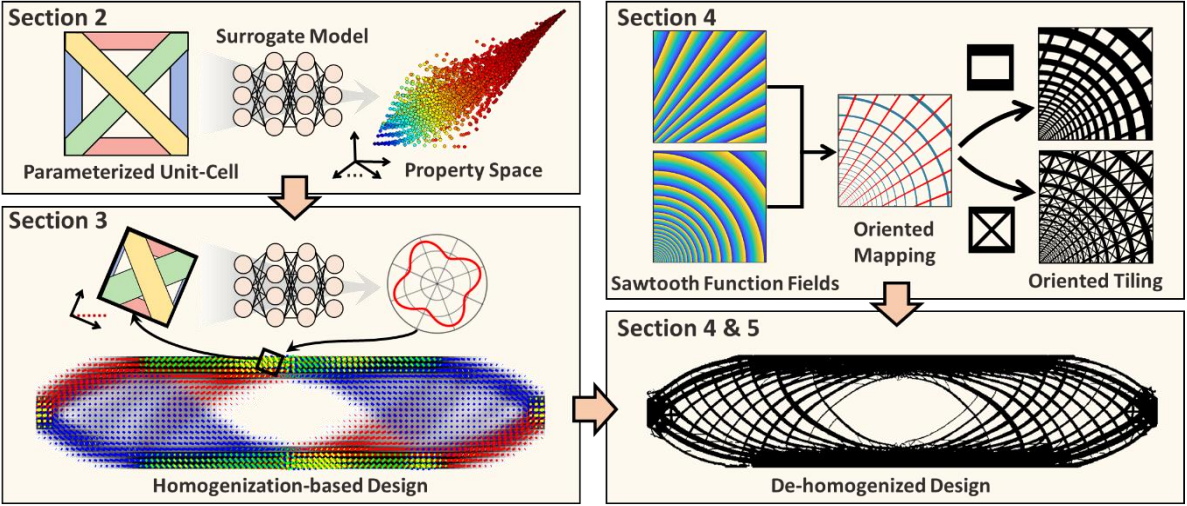

Fig. 1 Overview of the proposed framework.

## 2. Microstructure database construction and surrogate modeling

To meet the requirements for general design cases, diverse microstructures are desirable to cover a wide range of mechanical properties. Meanwhile, microstructures should be able to connect with each other to ensure boundary compatibility and manufacturability. Herein, we present a six-bar parameterized model, as shown in Fig. 2(a), to describe unit-cell microstructures in the multiscale design. Those bars are divided into four groups (indicated by different colors in Fig. 2(a)) based on their orientation, each assigned with an independent design parameter, i.e., $x_1$ through $x_4$. We choose this six-bar microscale representation because it offers a low-dimensional geometrical representation that can form a wide range of microstructure designs while ensuring boundary compatibility. To illustrate this, we show some representative microstructures generated by the six-bar model in Fig. 2(b). These microstructures have various shapes and topologies, and are guaranteed to have a compatible connection with each other. The flexibility in geometries brings diversity in properties (stiffness tensors), as demonstrated in the corresponding modulus surfaces in Fig. 2(c). By combining rods in different sizes and directions, one could obtain various symmetry types and directional characteristics of the microstructures. It should be emphasized that, although a six-bar representation is adopted here, the later modeling and optimization methods proposed in this study can also accommodate other rectangular cells with free-form geometries that are self-connected and properly parameterized.

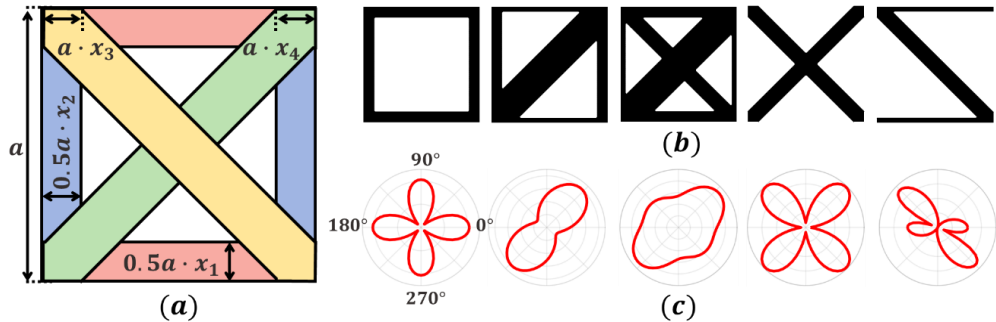

Fig. 2 Parameterization of the unit-cell microstructure, (a) Proposed six-bar model with bar groups colored differently, (b) representative microstructures and their (c) corresponding normalized modulus surfaces.

We generate $14^4 = 38416$ microstructures by sampling 14 width values, i.e., $x_i$, for each group of bars. These microstructures are discretized into a $100 \times 100$ pixel matrix, with zero and one to represent void and solid, respectively. Effective stiffness tensors of these microstructures are then calculated via energy-based homogenization. The constituent material has Young's modulus $E_0 = 210$ GPa and Poisson's ratio $v_0 = 0.3$. The property space spanned by independent entries of the stiffness tensor (in Voigt notation) and volume fraction, i.e., $E_{11}, E_{12}, E_{13}, E_{22}, E_{23}, E_{33}$ and $\rho$, is shown in Fig. 3. From the figure, it can be noted that the constructed microstructure database covers a wide range of mechanical properties, which will benefit our later multiscale optimization with better macroscale performance.

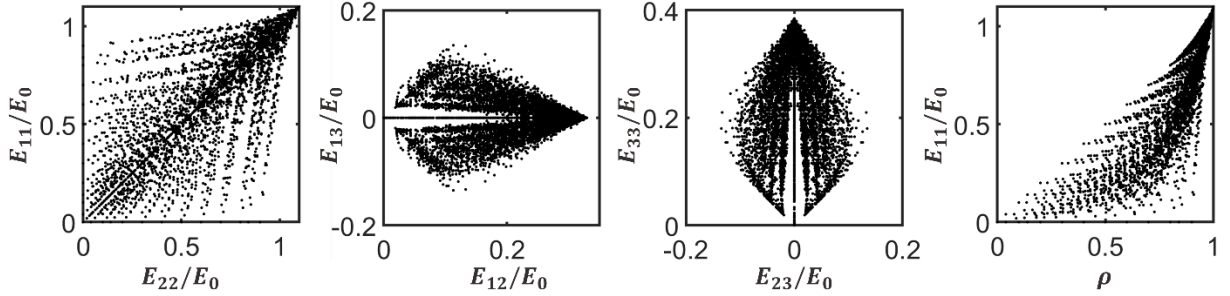

Fig. 3 Property space of the generated microstructure database, independent entries of the stiffness tensor are normalized with respect to the Young's modulus of the constituent material.

Since numerous on-the-fly homogenizations cause an unaffordable computational burden, a surrogate model is needed to efficiently approximate the relation between microscale geometrical parameters ($x_1, x_2, x_3, x_4$) and corresponding effective properties ($E_{11}, E_{12}, E_{13}, E_{22}, E_{23}, E_{33}$ and $\rho$). Herein, we opt for a multi-layer neural network due to its excellent capability in handling nonlinear relations and large datasets. As shown in Appendix A, the neural network is composed of four hidden layers with tanh activation functions, with a gradient vector that is continuous and can be obtained analytically. We use the scaled conjugate gradient algorithm to train the model on the constructed database divided into training (80%) and testing (20%) sets, with mean square error as the loss function. After training, the model achieves a high predictive power, with R-squared values larger than 0.9999 on both training and testing datasets.

## 3. Multiscale topology optimization with oriented microstructures

**3.1 Design representation**

In this section, we integrate the trained neural network into the multiscale topology optimization for frequency response. Specifically, as shown in Fig. 4, we discretize the design region $\Omega$ into a mesh $\mathcal{H}_{ma}$ of four-node quadrilateral elements, each with equal size $h_{ma}$. For each element, we need to design both the microstructure geometry and its orientation.

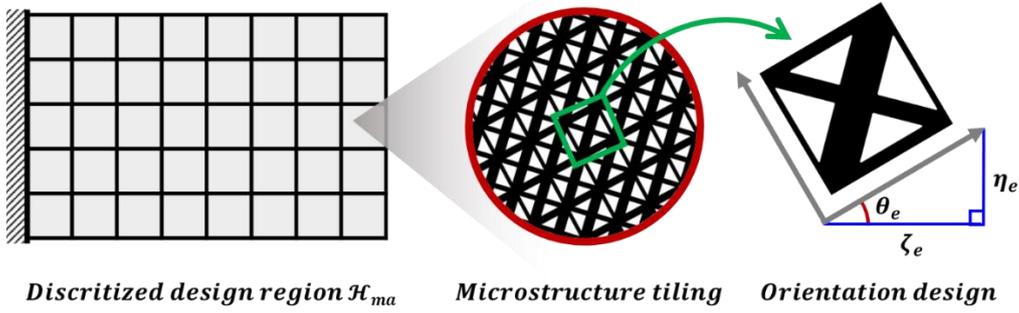

Fig. 4 Illustration of the multiscale design.

For the geometry design, the four parameters defined in Fig. 2 are used as the design variables for each element to describe its geometry, denoted as $x_e = [x_1^e, x_2^e, x_3^e, x_4^e]^T$. To avoid thin strips of solid or void in the microstructures, we will perform filtering on the geometrical design variables during the optimization via the projection scheme given as [28]

$$\tilde{x}_i^e = x_i^e \cdot \tilde{H}(x_i^e, \gamma, \tau) \cdot \left(1 - \tilde{H}(x_i^e, 1-\gamma, \tau)\right) + \left(\frac{\tau-1}{\tau} + \frac{x_i^e}{\tau}\right) \cdot \tilde{H}(x_i^e, 1-\gamma, \tau), \quad (1)$$

where $\tilde{H}$ is an approximated Heaviside function,

$$\tilde{H}(x_i^e, \gamma, \tau) = \frac{tanh(\tau\gamma) + tanh(\beta(x_i^e - \gamma))}{tanh(\tau\gamma) + tanh(\beta(1-\gamma))}. \quad (2)$$

$\beta$ and $\eta$ are parameters used to control the sharpness and threshold of the projection, respectively. We will follow the continuation scheme in [28] to gradually increase $\gamma$ and $\tau$ in every 30 iterations from 0 and 100 to 0.05 and 400, respectively. This filtering technique encourages the design variables to converge to 0, 1, or within the range of $[\gamma, 1-\gamma]$. By feeding the projected elemental geometrical design vectors $\tilde{x}_e$ into the trained neural network, we can obtain the predicted volume fraction $\hat{\rho}_e$ and independent entries in the stiffness tensor, i.e., $\hat{E}_{11}^e, \hat{E}_{12}^e, \hat{E}_{13}^e, \hat{E}_{22}^e, \hat{E}_{23}^e, \hat{E}_{33}^e$, together with their partial derivatives. Those independent entries are used to assemble the elemental stiffness tensor $\widehat{\boldsymbol{E}}_e$ (in Voigt notation).

For the design of the microstructural orientation, the most straightforward method is to use the orientation angle $\theta_e$ as the design variable. Despite its easy implementation, this method is strongly sensitive to the initial design. It also suffers from ambiguity due to the $2\pi$ periodicity of $\theta_e$, which is a critical issue for the commonly used filtering techniques in topology optimization. Therefore, instead of using a polar representation, we opt for a Cartesian coordinate representation by introducing a 2D orientation design vector $[\zeta_e, \eta_e]^T \in [0,1]^2$, as shown in Fig. 4. The cosine and sine function of the orientation angle $\theta_e$ can be given as

$$\cos(\theta_e) = \tilde{\zeta}_e = \frac{\zeta_e}{\sqrt{\zeta_e^2 + \eta_e^2} + \varepsilon}, \sin(\theta_e) = \tilde{\eta}_e = \frac{\eta_e}{\sqrt{\zeta_e^2 + \eta_e^2} + \varepsilon}, \tag{3}$$

where $\varepsilon$ is a small constant $(10^{-6})$ to avoid singularities. In this way, the original orientation angle field is decomposed into two scalar fields, which allows a continuous transition between different angles and can avoid the $2\pi$ ambiguity problem. The rotated stiffness tensor (in Voigt notation) corresponding to the orientation angle $\theta_e$ can then be obtained by

$$\widetilde{\boldsymbol{E}}_e(\boldsymbol{x}_e, \zeta_e, \eta_e) = \boldsymbol{T}_e^T(\zeta_e, \eta_e)\widehat{\boldsymbol{E}}_e(\boldsymbol{x}_e)\boldsymbol{T}_e(\zeta_e, \eta_e), \tag{4}$$

where $\boldsymbol{T}_e$ is the usual transformation matrix and given as

$$\boldsymbol{T}_e(\zeta_e, \eta_e) = \begin{bmatrix} \tilde{\zeta}_e^2 & \tilde{\eta}_e^2 & \tilde{\zeta}_e\tilde{\eta}_e \\ \tilde{\eta}_e^2 & \tilde{\zeta}_e^2 & -\tilde{\zeta}_e\tilde{\eta}_e \\ -2\tilde{\zeta}_e\tilde{\eta}_e & 2\tilde{\zeta}_e\tilde{\eta}_e & \tilde{\zeta}_e^2 - \tilde{\eta}_e^2 \end{bmatrix}. \tag{5}$$

### 3.2 Topology optimization problem definition

With the proposed geometry and orientation representations, we can obtain the corresponding $\widetilde{\boldsymbol{E}}_e$ and $\hat{\rho}_e$ to assemble the elemental stiffness matrix $\boldsymbol{k}_e$ and the elemental mass matrix $\boldsymbol{m}_e$

$$\boldsymbol{k}_e\left(\widetilde{\boldsymbol{E}}_e(\boldsymbol{x}_e, \zeta_e, \eta_e)\right) = \int_{\Omega_e} \boldsymbol{B}^T \widetilde{\boldsymbol{E}}_e \boldsymbol{B} d\Omega, \tag{6}$$

$$\boldsymbol{m}_e(\hat{\rho}_e(\boldsymbol{x}_e)) = \rho_0 \cdot \int_{\Omega_e} \hat{\rho}_e \boldsymbol{N}^T \boldsymbol{N} d\Omega, \tag{7}$$

where $\Omega_e$ is the region of the element, $N$ is the shape function matrix for the four-node element, $B$ is the associated constant gradient matrix (also known as strain-displacement matrix), and $\rho_0$ is the mass density of the constituent material. The momentum equation of a damped structure system under external harmonic forces is thus given as

$$M\ddot{\widetilde{U}} + C\dot{\widetilde{U}} + K\widetilde{U} = \widetilde{F}, \tag{8}$$

where $M, C$ and $K$ are $2n_e \times 2n_e$ global mass, damping, and stiffness matrices for the $n_e$ finite elements in the structure, respectively. $\widetilde{U}$ and $\widetilde{F}$ are $2n_e \times 1$ displacement vector and external force vector, respectively. $K$ and $M$ can be assembled by elemental stiffness matrix $k_e$ and the elemental mass matrix $m_e$, respectively. In this study, we assume a Rayleigh damping for the structure, so that the damping matrix $C$ can be expressed as

$$C = \alpha_r M + \beta_r K, \tag{9}$$

where $\alpha_r$ and $\beta_r$ are real scalar constants. Under the harmonic excitation assumption, we could further obtain

$$\widetilde{F} = Fe^{i\omega t}, \widetilde{U} = Ue^{i\omega t} \tag{10}$$

where $i$ is the imaginary unit, $\omega$ is the angular frequency of the excitation, $F$ and $U$ are the amplitudes of the external force and the displacement, respectively. With (10), we can rewrite (6) to be

$$SU = F, \tag{11}$$

$$S = -\omega^2 M + i\omega C + K, \tag{12}$$

where $S$ is known as the dynamic stiffness matrix.

The objective for frequency response optimization is to minimize the amplitude of the target response $J$ at a specific point or region of the structure, over a given frequency range $[\omega_l, \omega_u]$ of the excitation force. While various forms of objective functions have been proposed, we will use the following definition of the target response $J$ to measure the local displacement response [46]

$$J(x_e, \zeta_e, \eta_e | \omega_l, \omega_u) = \int_{\omega_l}^{\omega_u} \varphi(x_e, \zeta_e, \eta_e | \omega) d\omega = \int_{\omega_l}^{\omega_u} \sqrt{|U^T L \overline{U}|} d\omega, \tag{13}$$

where $\varphi$ is the response value under a given frequency $\omega$ of the excitation force, $\overline{U}$ is the conjugate of $U$, $|\cdot|$ is an operator to calculate the modulus of a complex number, and $L$ is an $2n_e \times 2n_e$ matrix with the value 1 at the degrees of freedom in the diagonal line corresponding to the displacements of interest, and with zeros at all other entries. In practical implementation, this objective function is usually approximated by various numerical quadrature techniques. Herein, for simplicity, we follow [45] to divide the frequency range into $n_s \geq 1$ small equal subintervals, and then use the trapezoidal summation to approximate the integration

$$J \cong \hat{J}(x_e, \zeta_e, \eta_e | \omega_l, \omega_u) = \sum_{i=1}^{n_s} \frac{\Delta\omega}{2} \left( \varphi(x_e, \zeta_e, \eta_e | \omega_i) + \varphi(x_e, \zeta_e, \eta_e | \omega_{i+1}) \right), \quad (14)$$

$$\omega_i = \omega_l + \frac{i-1}{n_s} \Delta\omega, \quad (15)$$

$$\Delta\omega = \frac{\omega_u - \omega_l}{n_s}, \quad (16)$$

where $\Delta\omega$ is the length of subintervals and $\omega_i$ is the endpoint of the subinterval. We can then define the topology optimization problem as

$$\min_{x_e, \zeta_e, \eta_e} \hat{J}(x_e, \zeta_e, \eta_e | \omega_l, \omega_u)$$
$$s.t. \ \boldsymbol{S}(\omega, x_e, \zeta_e, \eta_e) \boldsymbol{U} = \boldsymbol{F},$$
$$V_* - \frac{1}{n_e} \sum_{e=1}^{n_e} \hat{\rho}_e(x_e) \leq 0, \quad (17)$$
$$0 < \zeta_e, \eta_e \leq 1,$$
$$0 < x_{min} \leq x_i^e \leq 1, \quad i = 1, 2, \dots, 3,$$

where $V_*$ is the target solid material volume fraction of the full structure, $x_{min}$ is a small value ($10^{-6}$) to avoid singularities.

### 3.3 Sensitivity analysis

To solve the topology optimization problem defined in (17), sensitivity analysis is required to enable the use of an efficient gradient-based solver, such as the method of moving asymptotes (MMA) [54]. In this section, we use the adjoint method to derive the analytical sensitivity of the objective function. Note that the objective function can be viewed as a weighted sum of $\varphi(x_e, \zeta_e, \eta_e | \omega)$ for a set of discrete frequency values $\omega_i$. We can first derive the sensitivity of $\varphi$ for different $\omega$ values and then perform a weighted sum to obtain the sensitivity of $\hat{J}$. To achieve this, an augmented function is introduced for the original formula of $\varphi$:

$$\mathcal{F}(x_e, \zeta_e, \eta_e | \omega) = \varphi(x_e, \zeta_e, \eta_e | \omega) + \boldsymbol{\lambda}_1^T(\omega)(\boldsymbol{SU} - \boldsymbol{F}) + \boldsymbol{\lambda}_2^T(\omega)(\overline{\boldsymbol{S}}\overline{\boldsymbol{U}} - \overline{\boldsymbol{F}}), \quad (18)$$

where $\boldsymbol{\lambda}_1$ and $\boldsymbol{\lambda}_2$ are arbitrary vectors serving as the multiplier. Taking the derivative with respect to design variables (taking geometrical design variable $x_i^e$ as an instance) on both sides of the equation (18), we obtain

$$\frac{\partial \mathcal{F}}{\partial x_i^e} = \left(\frac{\overline{\boldsymbol{U}}^T \boldsymbol{L}}{2\varphi} + \boldsymbol{\lambda}_1^T \boldsymbol{S}\right)\frac{\partial \boldsymbol{U}}{\partial x_i^e} + \left(\frac{\boldsymbol{U}^T \boldsymbol{L}}{2\varphi} + \boldsymbol{\lambda}_2^T \overline{\boldsymbol{S}}\right)\frac{\partial \overline{\boldsymbol{U}}}{\partial x_i^e} + \boldsymbol{\lambda}_1^T \frac{\partial \boldsymbol{S}}{\partial x_i^e}\boldsymbol{U} + \boldsymbol{\lambda}_2^T \frac{\partial \overline{\boldsymbol{S}}}{\partial x_i^e}\overline{\boldsymbol{U}}, \qquad (19)$$

This can be transformed into

$$\frac{\partial \mathcal{F}}{\partial x_i^e} = \boldsymbol{\lambda}_1^T \frac{\partial \boldsymbol{S}}{\partial x_i^e}\boldsymbol{U} + \boldsymbol{\lambda}_2^T \frac{\partial \overline{\boldsymbol{S}}}{\partial x_i^e}\overline{\boldsymbol{U}}, \qquad (20)$$

when $\boldsymbol{\lambda}_1$ and $\boldsymbol{\lambda}_2$ satisfy the following equations:

$$\overline{\boldsymbol{U}}^T \boldsymbol{L} + 2\varphi \boldsymbol{\lambda}_1^T \boldsymbol{S} = \boldsymbol{0}, \boldsymbol{U}^T \boldsymbol{L} + 2\varphi \boldsymbol{\lambda}_2^T \overline{\boldsymbol{S}} = 0 \qquad (21)$$

Since we have $\boldsymbol{S}^T = \boldsymbol{S}$ and real-valued $\varphi$ and $\boldsymbol{L}$, the solution for these two equations can be obtained as

$$\boldsymbol{\lambda}_1^T = -\frac{1}{2\varphi}\overline{\boldsymbol{U}}^T \boldsymbol{L} \boldsymbol{S}^{-1}, \boldsymbol{\lambda}_2 = \overline{\boldsymbol{\lambda}}_1 \qquad (22)$$

After substituting (22) into (20), we can obtain

$$\frac{\partial \varphi}{\partial x_i^e} = \frac{\partial \mathcal{F}}{\partial x_i^e} = -\frac{1}{\varphi} Re\left(\overline{\boldsymbol{U}}^T \boldsymbol{L} \boldsymbol{S}^{-1} \frac{\partial \boldsymbol{S}}{\partial x_i^e}\boldsymbol{U}\right), \qquad (23)$$

where $Re(\cdot)$ selects the real part of the complex value, and $\frac{\partial \boldsymbol{S}}{\partial x_i^e}$ can be obtained by taking the derivative of (12)

$$\frac{\partial \boldsymbol{S}}{\partial x_i^e} = -\omega^2 \frac{\partial \boldsymbol{M}}{\partial \hat{\rho}_e}\frac{\partial \hat{\rho}_e}{\partial x_i^e} + i\omega \frac{\partial \boldsymbol{C}}{\partial \hat{\rho}_e} + \sum_{j,k=1}^{3}\frac{\partial \boldsymbol{K}}{\partial \tilde{E}_{e,jk}}\frac{\partial \tilde{E}_{e,jk}}{\partial x_i^e}. \qquad (24)$$

Substituting (9) into (24), we can obtain

$$\frac{\partial \boldsymbol{S}}{\partial x_i^e} = (-\omega^2 + i\omega\alpha_r)\frac{\partial \boldsymbol{M}}{\partial \hat{\rho}_e}\frac{\partial \hat{\rho}_e}{\partial x_i^e} + (-\omega^2 + i\omega\beta_r)\cdot \sum_{j,k=1}^{3}\frac{\partial \boldsymbol{K}}{\partial \tilde{E}_{e,jk}}\frac{\partial \tilde{E}_{e,jk}}{\partial x_i^e}, \qquad (25)$$

where $\frac{\partial \hat{\rho}_e}{x_i^e}$ and $\frac{\partial \tilde{E}_{e,jk}}{x_i^e}$ can be readily obtained via the trained neural network. The sensitivity of the original objective function $\frac{\partial \hat{J}}{\partial x_i^e}$ can then be formulated as

$$\frac{\partial \hat{J}}{\partial x_i^e} = \sum_{i=1}^{n_s}\frac{\Delta\omega}{2}\left(\frac{\partial \varphi}{\partial x_i^e}(x_e,\zeta_e,\eta_e|\omega_i) + \frac{\partial \varphi}{\partial x_i^e}(x_e,\zeta_e,\eta_e|\omega_{i+1})\right), \qquad (26)$$

The sensitivity of $\hat{J}$ with respect to the orientation design parameter $\zeta_e$ and $\eta_e$ can be obtained by simply replacing $\partial x_i^e$ with $\partial \zeta_e$ and $\partial \eta_e$, respectively, in both equations (25) and (26). Since $\hat{\rho}_e$ does not depend on $\zeta_e$ and $\eta_e$, we could simplify $\frac{\partial \boldsymbol{S}}{\partial \zeta_e}$ and $\frac{\partial \boldsymbol{S}}{\partial \eta_e}$ to be

$$\frac{\partial \boldsymbol{S}}{\partial \zeta_e} = (-\omega^2 + i\omega\beta_r)\cdot \sum_{j,k=1}^{3}\frac{\partial \boldsymbol{K}}{\partial \tilde{E}_{e,jk}}\frac{\partial \tilde{E}_{e,jk}}{\partial \zeta_e}, \qquad (27)$$

$$\frac{\partial \mathbf{S}}{\partial \eta_e} = (-\omega^2 + i\omega\beta_r) \cdot \sum_{j,k=1}^{3} \frac{\partial \mathbf{K}}{\partial \tilde{E}_{e,jk}} \frac{\partial \tilde{E}_{e,jk}}{\partial \eta_e}. \tag{28}$$

We can then employ MMA to solve the optimization problem (17) iteratively based on the sensitivity analysis. The sensitivity filter and PDE filter are implemented for geometrical design variables and orientation variables, respectively, to avoid numerical instability and possible defects [55].

## 4. Microstructure tiling with generalized de-homogenization

A common drawback of most existing data-driven multiscale TO methods is that they do not consider the design of unit-cell orientation [19-21]. The difficulties lie in that it is challenging to assemble microstructures in a way that can meet the designed orientation field and preserve the effective properties. De-homogenization opens a new venue to achieve this oriented microstructure tiling, but it was originally proposed for square cells with rectangular holes instead of free-formed microstructures. Therefore, existing designs via de-homogenization are confined to simple static compliance design. In this section, we propose to integrate a Sawtooth-function-based mapping into the original de-homogenization method [27, 28]. We will show that by incorporating this new mapping method, de-homogenization can readily realize oriented tiling of complex microstructures without much change to the original framework [28], so that it can handle general design cases, including both static and dynamic problems.

**4.1 Sawtooth function**

The key of the proposed generalized de-homogenization method is to replace the original cosine function with a sawtooth function, which is commonly used in signal processing [41]. Specifically, for $y \in R$, the sawtooth function $\psi(y)$ can be formulated as

$$\psi(y) = -\frac{2}{\pi} \cdot arctan\left(cot\left(\frac{\pi}{P}y\right)\right), \tag{29}$$

where $P$ is a parameter to control the periodicity. As shown in Fig. 5 (a), $\psi$ stars from -1, linearly increases to 1 and then suddenly bounces back to 0, repeating this pattern with a periodicity of $P$. The corresponding response curve has a sawtooth-like shape. We can generalize this function to 2D cases by defining

$$\psi(\boldsymbol{y}) = -\frac{2}{\pi} \cdot arctan\left(cot\left(\frac{\pi}{P}\boldsymbol{e}(\theta)^T\boldsymbol{y}\right)\right), \tag{30}$$

where $\boldsymbol{y} = [y_1, y_2] \in \boldsymbol{R}^2$ is a 2D position vector and $\boldsymbol{e}(\theta)$ is a unit vector with directional angle $\theta$. We call $\boldsymbol{e}(\theta)$ a directional vector since it guides the propagation of the 2D sawtooth function. As shown in Fig. 5(b) and (c), we can easily control the propagation direction of the 2D sawtooth field by changing the directional vector. It should be noted that the directional vector can be spatially varying, i.e., a function of $\boldsymbol{y}$, instead of a constant vector. For example, as demonstrated in Fig. 5(d), we realize a complex field with spatially varying propagation directions by assigning the normalized position vector as the directional vector of the sawtooth function. Meanwhile, we can also control the periodicity of the sawtooth field by changing $P$ (compare Fig. 5(b) with Fig. 5(e) and (f)). Similarly, $P$ can also be a function of $\boldsymbol{y}$ to obtain spatially varying periodicity, as shown in Fig. 5(g). An inherent advantage of this field-based representation is that it allows a smooth transition between different patterns by interpolating $P$ and $\boldsymbol{e}(\theta)$, as demonstrated in Fig. 5(h) and (i). Therefore, this sawtooth function provides a simple and yet very flexible method to represent oriented fields, which will benefit the later microstructure tiling.

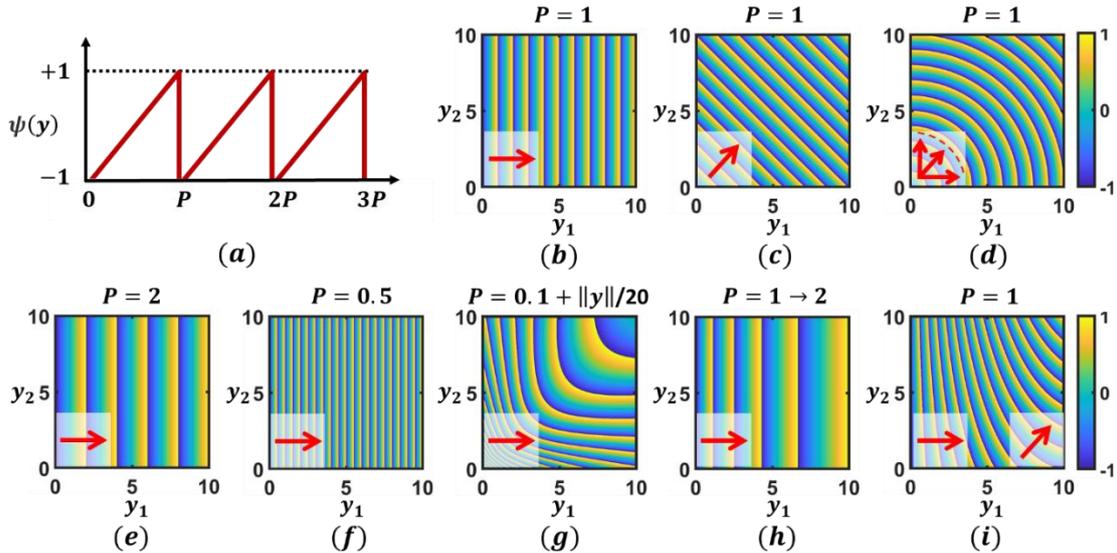

Fig. 5 Instances for 1D and 2D sawtooth function field, (a) a 1D sawtooth function field, (b)~(d) 2D function fields with the same $P = 1$ but with different $\theta$s, $\boldsymbol{e}(\theta)$ of each field is marked

in the lower-left inset, (e)~(f) 2D function fields with the same $\theta = 0$, but with different $P$s, (h) 2D function fields with the same $\theta = 0$, $P$ gradually transforms from 1 to 2 as $y_1$ increases, (i) 2D function fields with the same $P = 1$, $\theta$ gradually transforms from 0 to $\pi/4$ as $y_1$ increases.

### 4.2 Sawtooth-function-based microstructure tiling

We view microstructure tiling as a texture mapping that locally aligns with a given orientation field. To achieve this mapping, we first generate an oriented mesh that follows the orientation vectors by combining two 2D sawtooth function fields $\psi_1(y)$ and $\psi_2(y)$. This oriented mesh can be used to map unit-cell designs to corresponding elements in generating target microstructure tiling. In this subsection, we will focus on the constant orientation field to illustrate this sawtooth-function-based microstructure mapping. The mapping will be generalized to spatially varying orientation fields to enable de-homogenization in the next subsection.

Given a constant orientation field $e(\theta)$ and spatially varying periodicity field $P(y)$, we can define $\psi_1(y)$ and its orthogonal counterpart $\psi_2(y)$ as

$$\psi_1(y) = -\frac{2}{\pi} \cdot arctan\left(cot\left(\frac{\pi}{P(y)} e_1^T y\right)\right) = -\frac{2}{\pi} \cdot arctan\left(cot\left(\frac{\pi}{P(y)} e(\theta)^T y\right)\right), \quad (31)$$

$$\psi_2(y) = -\frac{2}{\pi} \cdot arctan\left(cot\left(\frac{\pi}{P(y)} e_2^T y\right)\right) = -\frac{2}{\pi} \cdot arctan\left(cot\left(\frac{\pi}{P(y)} e\left(\theta + \frac{\pi}{2}\right)^T y\right)\right). (32)$$

These two fields form a dyadic $\boldsymbol{\psi} = (\psi_1, \psi_2)$, which can be viewed as a mapping $\boldsymbol{\psi}(y): R^2 \to [-1,1]^2$. As shown in Fig. 6, this dyadic representation or mapping $\boldsymbol{\psi}$ actually creates an oriented tiling mesh composed of quadrilateral regions $[-1,1]^2$ that locally align with the given constant direction field $e(\theta)$.

For a given unit-cell microstructure, we can associate the unit-cell region with a coordinate system $(t_1, t_2) \in [-1,1]^2$. While there are various shape descriptors to represent the unit-cell geometry, they are in essence different ways to parameterize the mapping $\chi(t_1, t_2)$ from the unit-cell region $(t_1, t_2)$ to a scalar field that indicates the distribution of solid and void. For example, a pixelated representation maps the unit-cell region into a binary scalar field with 0 and 1 representing solid and void. A level set representation maps the unit-cell region into a real value scalar field and uses a threshold to differentiate solids and voids. Therefore, we could perform the mapping operation $\chi(t_1, t_2)$ on the dyadic $\psi$ formed by the two sawtooth fields $\psi_1(y)$ and $\psi_2(y)$ to obtain a scalar field $\chi(\psi_1(y), \psi_2(y))$ that represents the corresponding microstructure tiling. As demonstrated in Fig. 7, this approach enables the oriented tilling for any quadrilateral cells with free-form geometries. The extension of this mapping method to 3D cases is straightforward by adding another sawtooth field perpendicular to the 2D plane.

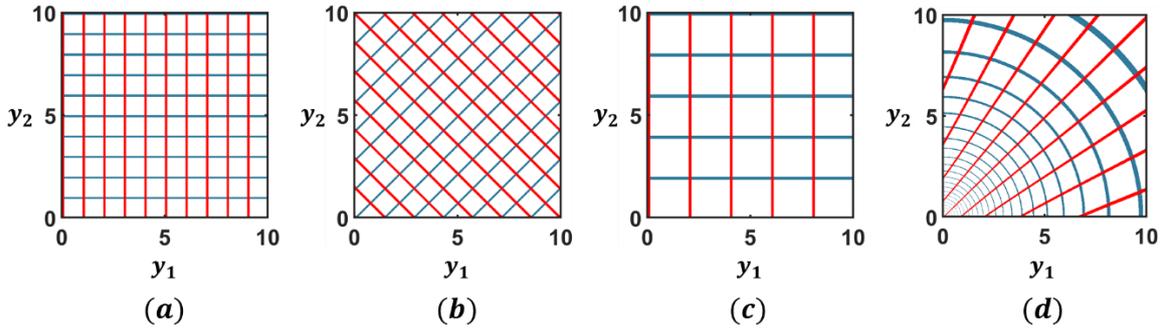

Fig. 6 Oriented tiling meshes generated by two sawtooth function fields. The contour lines for the peak values of $\psi_1(y)$ and $\psi_2(y)$ are colored in red and blue, respectively.

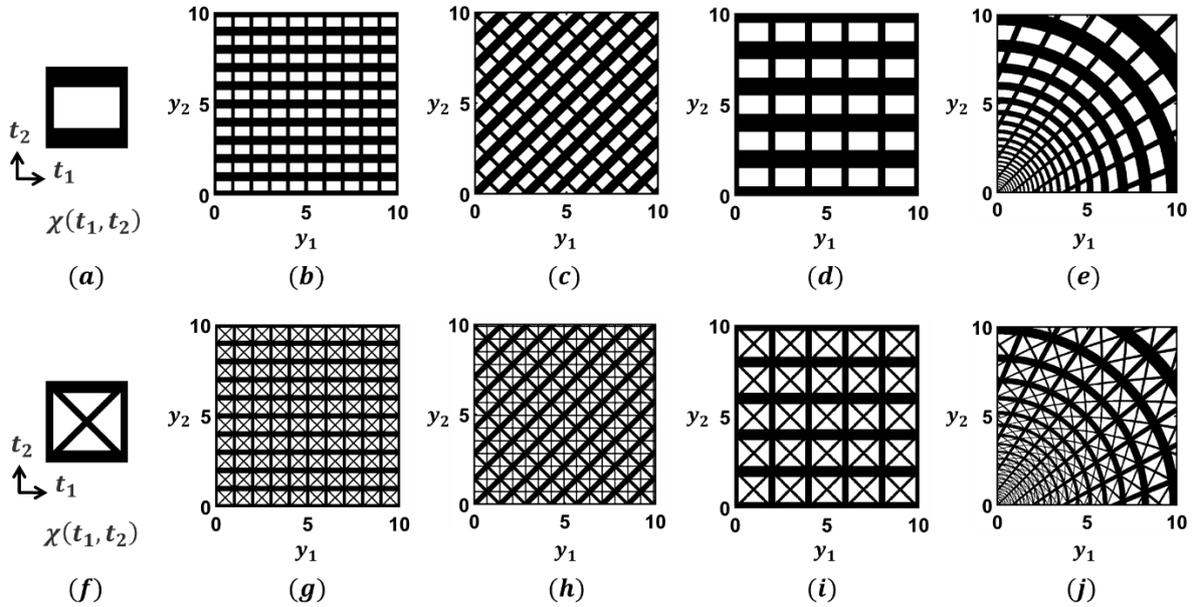

Fig. 7 Examples for oriented microstructure tiling, (a) unit-cell microstructure A and (b)~(e) its tiling corresponding to oriented mesh in Fig. 5, (f) unit-cell microstructure A and (g)~(j) its tiling corresponding to oriented mesh in Fig. 5.

Moreover, we can consider this tiling process as a two-stage composite mapping, with the first-stage mapping from $\mathbf{y}$ into $(t_1, t_2)$ via $\boldsymbol{\psi}(\mathbf{y})$ and the second-stage mapping from $(t_1(\mathbf{y}), t_2(\mathbf{y}))$ to $\chi$. As illustrated in Appendix B, it can be proved that the first-stage mapping is conformal (angle-preserving) within the unit cell, given that $P$ is a positive constant and the directional vectors $\mathbf{e_1}$ and $\mathbf{e_2}$ form an orthogonal pair. As a result, according to [5], the homogenized properties of the unit-cell microstructure are well-preserved in the mapped tiling. This is highly desirable in de-homogenization, and is the main reason we opt for sawtooth functions to construct the mapping.

### 4.3 Generalized de-homogenization for free-form microstructures

We propose to integrate sawtooth-function-based mapping into the de-homogenization framework and extend it to handle square cells with free-form geometries. In the previous subsection, we simply employ Equations (31) and (32) to generate oriented mesh that follows a given constant orientation field $\mathbf{e}(\theta)$ for microstructure mapping. However, as discussed in [38], it may not accommodate a spatially varying orientation field $\mathbf{e}(\theta(\mathbf{y}))$. To solve this issue, we relax the construction of the mapping into an optimization problem that finds the two sawtooth fields, $\psi_1(\mathbf{y})$ and $\psi_2(\mathbf{y})$, whose local propagation directions can best approximate the given orientation field $\mathbf{e_1} = \mathbf{e}(\theta(\mathbf{y}))$ and its orthotropic counterpart $\mathbf{e_2} = \mathbf{e}\left(\theta(\mathbf{y}) + \frac{\pi}{2}\right)$. Specifically, we introduce two scalar fields $\emptyset_1(\mathbf{y}) = \mathbf{e_1}^T \mathbf{y}$ and $\emptyset_2(\mathbf{y}) = \mathbf{e_2}^T \mathbf{y}$, instead of using $(\mathbf{e_1}, \mathbf{e_2})$, as the optimization variables to reduce the dimensionality. Taking the search for an optimized $\emptyset_1(\mathbf{y})$ as an example, we can formulate the optimization problem as

$$\min_{\emptyset_1(\mathbf{y})} \mathcal{G}_1\big(\emptyset_1(\mathbf{y})\big) = \frac{1}{2} \int_\Omega d_1(\mathbf{y}) \|\nabla \emptyset_1(\mathbf{y}) - \mathbf{e}(\theta(\mathbf{y}))\|_2^2 \, d\Omega$$

$$s.t. \; d_2(\mathbf{y}) \nabla \emptyset_1(\mathbf{y}) \cdot \mathbf{e}\left(\theta(\mathbf{y}) + \frac{\pi}{2}\right) = 0, \tag{33}$$

where

$$d_1(\mathbf{y}) = \begin{cases} 0.01 & if \; \mathbf{y} \in \Omega_v \\ 0.1 & if \; \mathbf{y} \in \Omega_s \\ 1 & if \; \mathbf{y} \in \Omega_l \end{cases}, \quad d_2(\mathbf{y}) = \begin{cases} 0 & if \; \mathbf{y} \in \Omega_v \\ 0 & if \; \mathbf{y} \in \Omega_s \\ 1 & if \; \mathbf{y} \in \Omega_l \end{cases}, \tag{34}$$

where $\Omega_v, \Omega_s$ and $\Omega_l$ represent regions with volume fraction values $\hat{\rho} = 0$, $\hat{\rho} = 1$ and $\hat{\rho} \in [0,1]$, respectively. Herein, the objective function $\mathcal{g}_1(\emptyset_1(\mathbf{y}))$ measures how well the local propagation direction of the generated sawtooth function field, i.e., $\nabla \emptyset_1(\mathbf{y})$, aligns with the given direction field $\mathbf{e}(\theta(\mathbf{y}))$. The constraint is added to enforce the mapped microstructures to have orthogonal boundaries. As illustrated in the last subsection, this can ensure that the conformal mapping better preserves its original homogenized properties. The orientation designs of solids $\Omega_s$ and voids $\Omega_v$ will not influence the local homogenized properties, but tend to be badly determined (disordered orientations) that can lead to large distortion of the mapped structure. Therefore, the two terms $d_1$ and $d_2$ are added to relax the projection in those regions to avoid large distortion induced by badly determined orientations.

The Lagrangian equations of the optimization problem (33) can be given as

$$\mathcal{L}_1\left(\emptyset_1(\mathbf{y}), \tilde{\lambda}_1(\mathbf{y})\right) = \mathcal{g}_1(\emptyset_1(\mathbf{y})) - \int_\Omega \tilde{\lambda}_1(\mathbf{y}) d_2(\mathbf{y}) \nabla \emptyset_1(\mathbf{y}) \cdot \mathbf{e}\left(\theta(\mathbf{y}) + \frac{\pi}{2}\right) d\Omega, \quad (35)$$

where $\tilde{\lambda}_1(\mathbf{y})$ is a Lagrange multiplier. Similarly, we can obtain the Lagrangian equation for the second sawtooth function as

$$\mathcal{L}_2\left(\emptyset_2(\mathbf{y}), \tilde{\lambda}_2(\mathbf{y})\right) = \mathcal{g}_2(\emptyset_2(\mathbf{y})) - \int_\Omega \tilde{\lambda}_1(\mathbf{y}) d_2(\mathbf{y}) \nabla \emptyset_1(\mathbf{y}) \cdot \mathbf{e}(\theta(\mathbf{y})) d\Omega, \quad (36)$$

where $\tilde{\lambda}_2(\mathbf{y})$ is a Lagrange multiplier and $\mathcal{g}_2(\emptyset_2(\mathbf{y}))$ is given as

$$\mathcal{g}_2(\emptyset_2(\mathbf{y})) = \frac{1}{2} \int_\Omega d_1(\mathbf{y}) \left\| \nabla \emptyset_2(\mathbf{y}) - \mathbf{e}\left(\theta(\mathbf{y}) + \frac{\pi}{2}\right) \right\|_2^2 d\Omega. \quad (37)$$

In practical implementations, while the homogenization-based topology optimization proposed in Section 3 is performed on the mesh $\mathcal{H}_{ma}$ with elements of size $h_{ma}$, these two optimization problems in Equations (35) and (36) are solved on a finer mesh $\mathcal{H}_{sa}$ with element size $h_{sa} < h_{ma}/3$ as suggested in [38]. Optimized topology optimization variables obtained on $\mathcal{H}_{ma}$ are projected to $\mathcal{H}_{sa}$ by interpolation. We denote the number of elements of $\mathcal{H}_{sa}$ as $n_{e,sa}$. The solutions for the corresponding discretized Lagrange equations on $\mathcal{H}_{sa}$ can be efficiently obtained by solving the following KKT systems,

$$\begin{bmatrix} D^T A_1 D & -D^T B_2^T A_2^T \\ A_2 B_2 D & 0 \end{bmatrix} \begin{bmatrix} \emptyset_1 \\ \tilde{\lambda}_1 \end{bmatrix} = \begin{bmatrix} D^T A_1 C_1 \\ 0 \end{bmatrix}, \quad (38)$$

$$\begin{bmatrix} D^T A_1 D & -D^T B_1^T A_2^T \\ A_2 B_1 D & 0 \end{bmatrix} \begin{bmatrix} \emptyset_2 \\ \tilde{\lambda}_2 \end{bmatrix} = \begin{bmatrix} D^T A_1 C_2 \\ 0 \end{bmatrix}, \quad (39)$$

where $D \in R^{2n_{e,sa} \times 2n_{e,sa}}$ is a finite difference matrix, $A_1, A_2 \in R^{2n_{e,sa} \times 2n_{e,sa}}$ are diagonal weighted matrices with diagonal entries to be $d_1$ and $d_2$ at the corresponding points in $\mathcal{H}_{sa}$, respectively, vectors $C_1, C_2 \in R^{2n_{e,sa} \times 1}$ are composed of $e(\theta(y))$ and $e\left(\theta(y) + \frac{\pi}{2}\right)$ at the corresponding points in $\mathcal{H}_{sa}$, respectively, $B_1, B_2 \in R^{2n_{e,sa} \times 2n_{e,sa}}$ are diagonal matrices with $C_1$ and $C_2$ as diagonal lines, respectively. Note that the optimized orientation angle distribution may contain sudden changes of quadrants, i.e., rotated by π, due to the lack of rotation polarity. This could lead to infeasible geometrical features in the microstructure tiling. Therefore, we follow Ref. [38] and use a connected component labeling algorithm to identify those sudden changes and modify the angles to ensure a consistent orientation field.

After solving Equations (38) and (39), the solutions $\emptyset_1$ and $\emptyset_2$ are smoothed by a density filter with a radius of $h_{sa}$ to avoid local high-frequency variations. The smoothed $\emptyset_1$ and $\emptyset_2$ are then projected to a finer mesh $\mathcal{H}_{mi}$ through interpolation with element size $h_{mi}$ satisfying $h_{mi} < h_{ma}/15$ and $h_{mi} \leq h_{sa}$. These projected $\emptyset_1$ and $\emptyset_2$ are then substituted into Equations (31) and (32) to generate the corresponding sawtooth function fields $\psi_1$ and $\psi_2$ on the fine mesh $\mathcal{H}_{mi}$, from which the oriented microstructure tiling can be readily obtained as $\chi(\psi_1, \psi_2)$. To eliminate possible defects, such as isolated pixels and checkerboard patterns, we use morphological operators in MATLAB to detect and fix those defects to ensure the feasibility of the de-homogenized structure. Note that the periodicity parameter $P$ is assigned as constant and can be tuned to control the size of unit-cell microstructures in the full design.

## 5. Design Case Study

To demonstrate the effectiveness and features of the proposed method, we apply it to three design cases in this section. For all examples, the constituent material is the same as the one used in Section 2 for the construction of the microstructure database, with Young's modulus $E_0 = 210$ Gpa, Poisson's ratio $v_0 = 0.3$, and density $\rho_0 = 2700 \ kg/m^3$. In the initial design, we assume the structure is filled by microstructures composed of horizontal and vertical bar groups only. These two groups of bars in the initial design have the same width, achieving a volume fraction equal to the target global volume fraction $V_*$. The principal strain direction calculated at the lower bound of the given frequency range is used as the initial guess for the orientation field. Optimization terminates when the change in design variables (normalized) is less than 0.01 between two consecutive iterations or when the number of iterations reaches

300. For the de-homogenization process, we set $h_{mi} = h_{ma}/20$ and $h_{sa} = h_{ma}/4$.

### 4.1 Pinned beam

As the first case study, we focus on the design of a $1.8\ m \times 0.4\ m$ pinned beam shown in Fig. 8. External pressure with an amplitude of $50000 N/m$ is imposed on the top and bottom of the beam. The optimization objective is to minimize the frequency response of those loaded regions (marked in blue) within a given excitation frequency range of $[0, 200]\ Hz$, under a 50% volume fraction constraint. The beam is divided into a $90 \times 20$ mesh $\mathcal{H}_{ma}$ to perform the proposed homogenization-based optimization. We use 21 integration points in calculating the objective functions with equal subintervals.

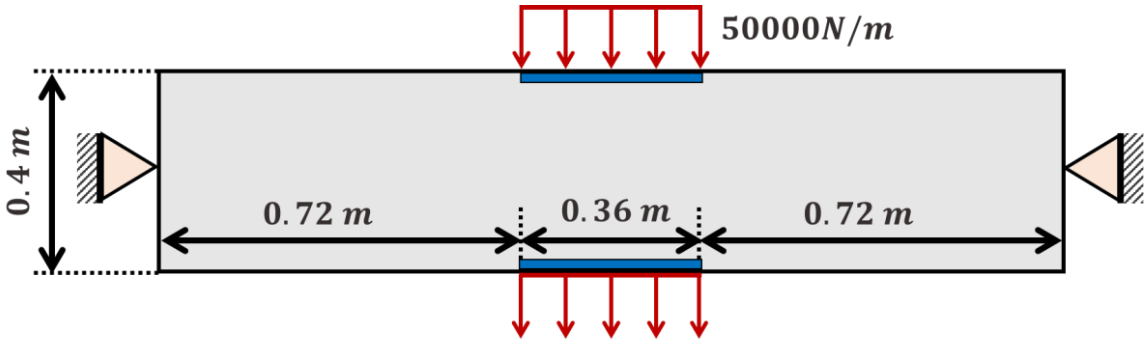

Fig. 8 Problem setting illustration of the first case study.

The optimized design variables distribution is demonstrated in Fig. 9(a) and (b). It can be noted that the orientation of microstructures largely aligns with the contour of the volume fractions, which strengthens the load-bearing capability of the structure. Different regions of the structure are dominated by different bar groups to better accommodate the spatially varying loading status. As a result, the optimized design demonstrates excellent dynamic performance, reducing the objective function value from $0.5193\ m \cdot Hz$ to only $0.0596\ m \cdot Hz$. As shown in Fig. 8 (c), the corresponding de-homogenized design, projected onto a $1800 \times 400$ fine mesh, is obtained via the proposed de-homogenization method with $P = 1.5$. Its oriented microstructure tiling matches well with the optimized orientation field shown in Fig. 9(b). The outer frame of the multiscale structure is composed of higher volume fraction microstructures conforming to the shape of the frame, while the inner region is filled by crossing bars to resist shearing forces. The de-homogenized design has a 50.25% volume fraction and achieves an objective value of $\hat{J} = 0.0637\ m \cdot Hz$, which is close to that of the homogenization-based design ($0.0596\ m \cdot Hz$). This highlights the effectiveness of the proposed de-homogenized method in generating a multiscale structure that realizes the designed performance of the homogenization-based design.

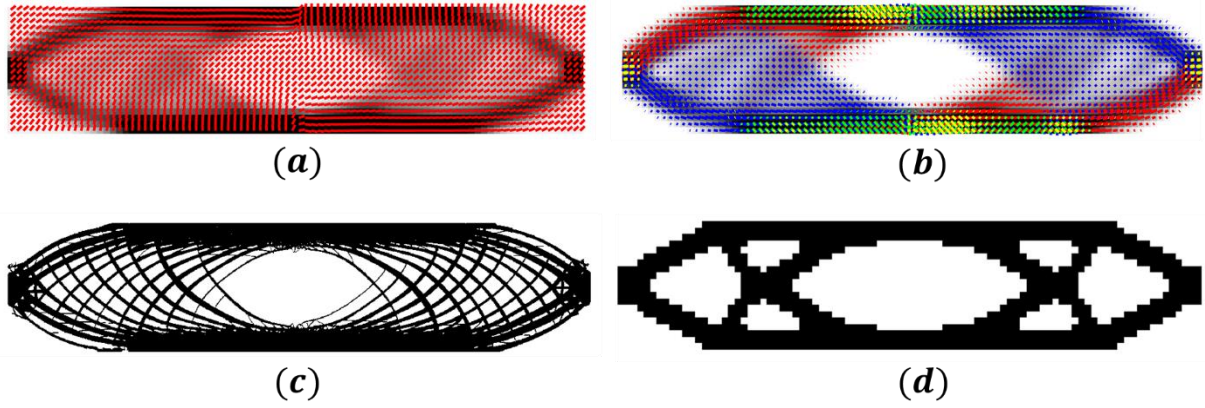

Fig. 9 Optimization results for the pinned beams, (a) designed volume fraction distribution with orientation vector marked by red line segments, (b) optimized design variables distribution, the widths of different bar groups are represented by the lengths of different line segments with the same color codes as in Fig. 1(a), these line segments are rotated by an angle corresponding to the designed orientation, (c) de-homogenized multiscale design inferred from (b), (d) single scale (macroscale) design.

To demonstrate the edge that the proposed multiscale design has over single-scale design, we use the classical SIMP method to design the beam under the same setting but only at the macroscale. The optimized structure is then projected onto the same fine mesh used in the de-homogenized design. The resultant macroscale design is shown in Fig. 9(d), with the objective value ($0.0705\ m \cdot Hz$) much higher than that of the de-homogenized design. This can also be observed from the frequency analysis shown in Fig. 10. Specifically, the initial design has a fundamental frequency $\omega_0 = 163.0349\ Hz$, which incurs a peak of the frequency response within the frequency range of interest. In contrast, the fundamental frequencies are much higher for both the de-homogenized (290.46 Hz) and macroscale design (287.24 Hz), moving the peaks of frequency responses away from the range of interest. As a result, the frequency responses of these two designs stay at a much lower level than the initial design. Moreover, the peak frequency response of the de-homogenized design has a lower amplitude, and corresponds to a higher frequency value, compared to the macroscale design. The response curve of the homogenized design stays below that of macroscale design over the whole frequency range of interest. This demonstrates the advantages of the proposed multiscale design over the macroscale design.

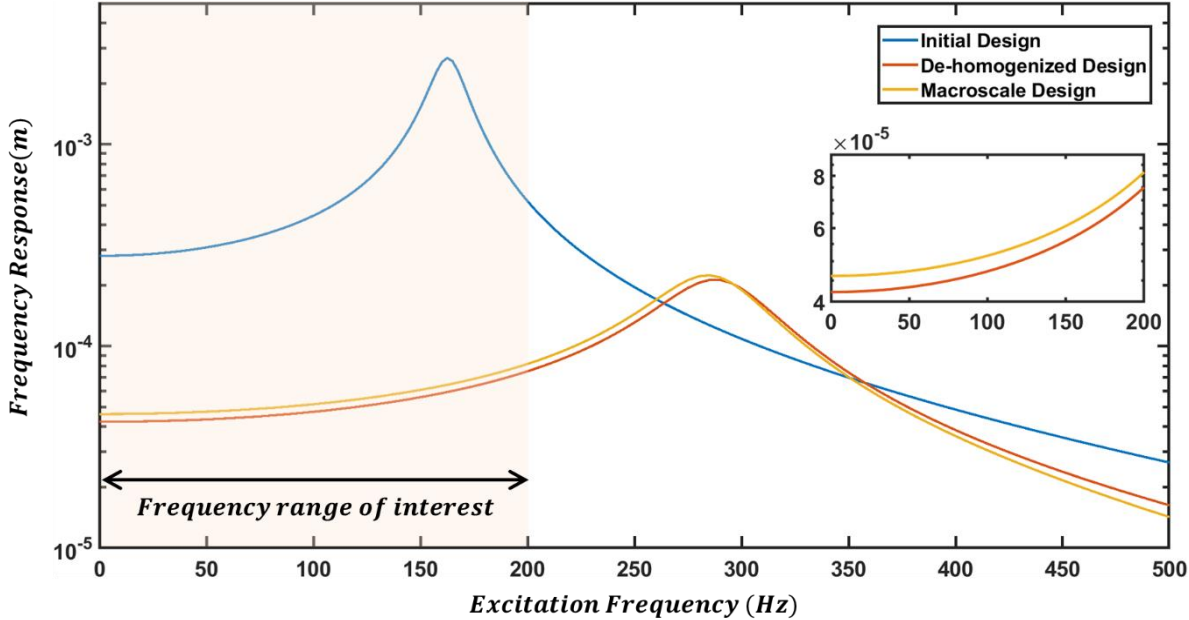

Fig. 10 Frequency response analysis for initial design, de-homogenized design, and macroscale design. The inlet shows an enlarged view of the frequency responses for de-homogenized design and macroscale design within the frequency range of interest.

In terms of efficiency, the proposed method takes 314.54s (300 iterations) to do homogenization-based optimization and another 349.96s to perform de-homogenization on a single CPU. While the total execution time of the proposed method is around 2 times longer than that of the macroscale SIMP method (332.51s), it achieves much better structural performance, as demonstrated earlier. In contrast, if classical SIMP is adopted to achieve the same resolution of the multiscale design in this study ($1800 \times 400 = 720000$ elements), the execution of a single iteration alone will take 3712.12s on the same platform, as the number of element and design variables is increased by 400 times compared with the design in the coarse mesh ($90 \times 20 = 1800$ elements). Considering the fact that a high-resolution design usually takes more iterations (~1000) to converge [28], the overall execution time for macroscale TO on fine meshes will become unaffordable (>1000 hours). Therefore, our proposed method shows a clear advantage in achieving excellent structural performance while retaining high efficiency (over 5700 times faster than TO on the fine mesh).

Moreover, as illustrated in Section 4.3, we can easily control the size of the unit-cell microstructures by changing the periodicity parameter $P$. For the same design variables distribution in Fig. 9 (b), we show the de-homogenized structures obtained with different $P$ values and their frequency responses in Figs. 11 and 12, respectively.

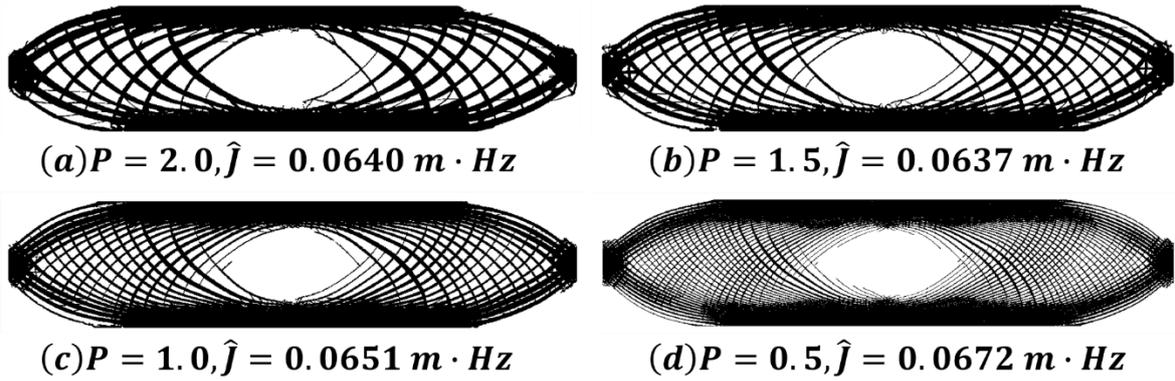

$(a) P = 2.0, \hat{J} = 0.0640 \ m \cdot Hz$    $(b) P = 1.5, \hat{J} = 0.0637 \ m \cdot Hz$

$(c) P = 1.0, \hat{J} = 0.0651 \ m \cdot Hz$    $(d) P = 0.5, \hat{J} = 0.0672 \ m \cdot Hz$

Fig. 11 De-homogenized designs obtained with different values of periodicity parameter $P$.

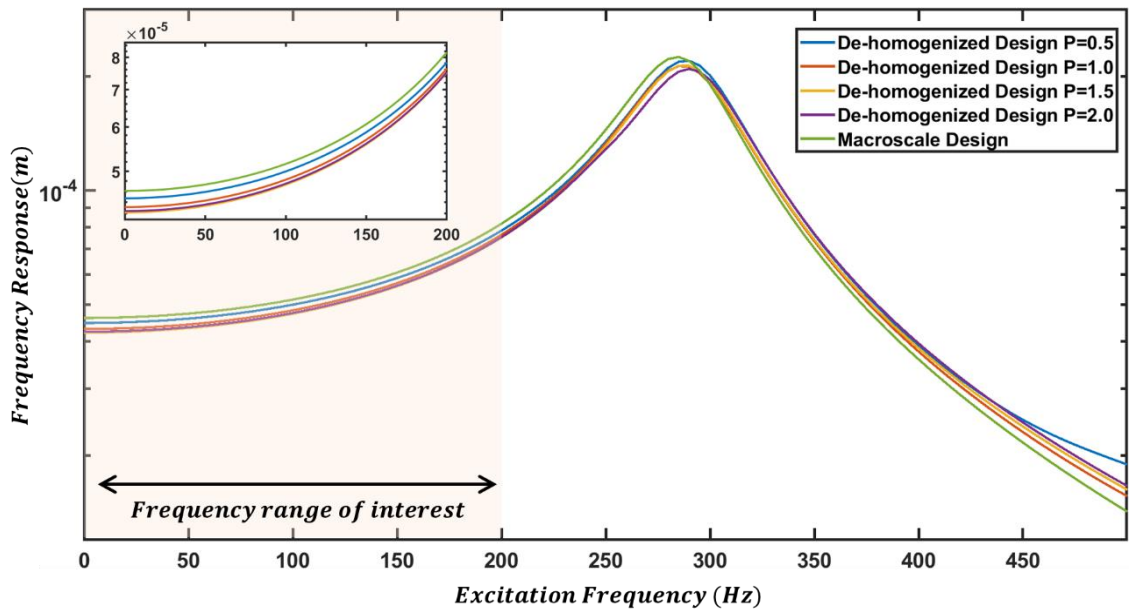

Fig. 12 Frequency response analysis for macroscale design and de-homogenized designs obtained with different values of periodicity parameter $P$. The inlet shows an enlarged view of the frequency responses within the range of interest.

From Fig. 11, it can be observed that the smaller the $P$ value, the smaller the size of unit-cell microstructures as well as the width of the bars. Despite the difference in topologies, all these structures exhibit similar unit-cell orientation and volume fraction distribution. As a result, these designs have similar objective values and frequency responses. Meanwhile, although the objective values should, in theory, decrease as the unit-cell size decreases and converge to that of homogenized design, such a pattern is not observed in our designs. The design with medium-size unit-cells ($P = 1.5$) has the best performance while the one with the smallest unit-cell size ($P = 0.5$) performs the worst. This is likely due to the fact that, given a fixed and finite resolution of the fine mesh, the bars of small-size cells would become so thin that they weaken their capability to resist the loads. Still, all de-homogenized designs have better performance than the single-scale macroscale design, regardless of the unit-cell size.

We continue to study how the unit-cell geometry parameterization method and orientation design will influence structural performance. Specifically, we align the microstructures along with the horizontal axis and fix the orientation during the optimization process. As shown in Figs. 13(a) and (b), we obtain two designs composed of microstructures with four bars (only horizontal and vertical bars in Fig. 2) and all six bars, respectively. Unlike the multiscale designs in Fig. 12, microstructures in the four-bar design fail to match with the macroscale geometry due to the lack of orientation design. As a result, the objective function ($0.1092\ m \cdot Hz$) is much higher than the oriented designs in Fig. 11, which is also indicated by the frequency analysis in Fig. 13(c). By including diagonal bars to obtain the six-bar design, the main load-bearing directions of micro-and macrostructures become more consistent, resulting in a lower objective function ($0.0890\ m \cdot Hz$). Nevertheless, its performance is far beneath the oriented de-homogenized designs in Fig. 11, underlining the importance of including diverse microstructures and orientation design for higher flexibility.

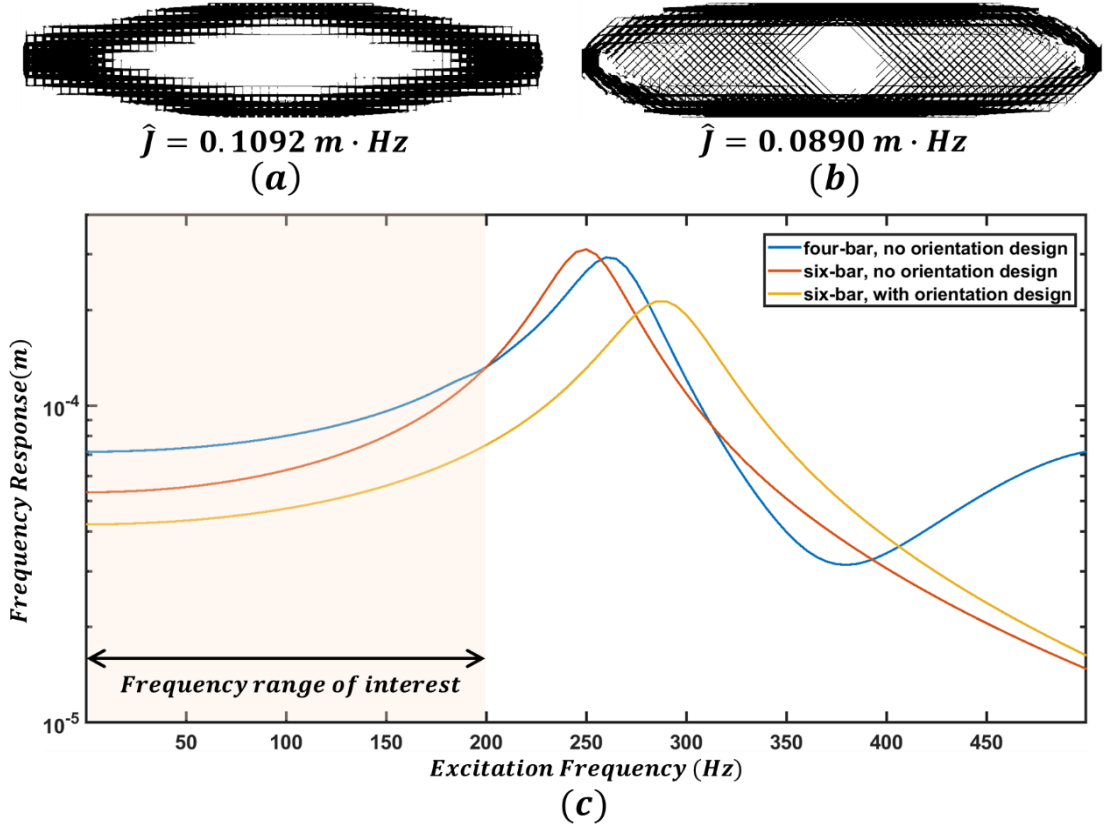

Fig. 13 Multiscale designs (a) with four bars but no orientation design, (b) with six bars but no orientation design, and (c) corresponding frequency response analysis.

**4.2 Clamped beam**

In this case study, as shown in Fig. 14, the proposed method is applied to design a clamped beam under a 50% volume fraction constraint, with the same loading and regions of interest as in the first case study. The beam is divided into a $90 \times 20$ mesh $\mathcal{H}_{ma}$ to perform the proposed homogenization-based optimization. We design the beam within different frequency ranges of interest, i.e., $[0,200]Hz, [100,300]Hz$ and $[200,400]Hz$. We use 21 integration points in calculating the objective functions with equal subintervals. With the periodicity parameter $P = 1.5$, the corresponding de-homogenized designs are shown in Fig. 15.

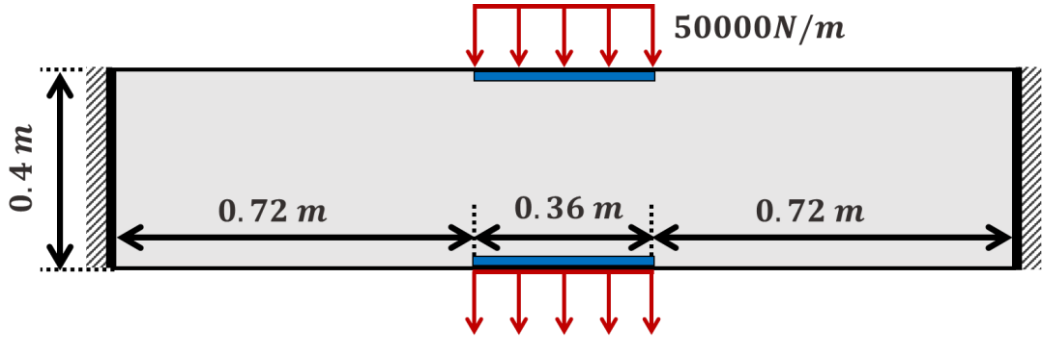

Fig. 14 Problem setting illustration of the second example.

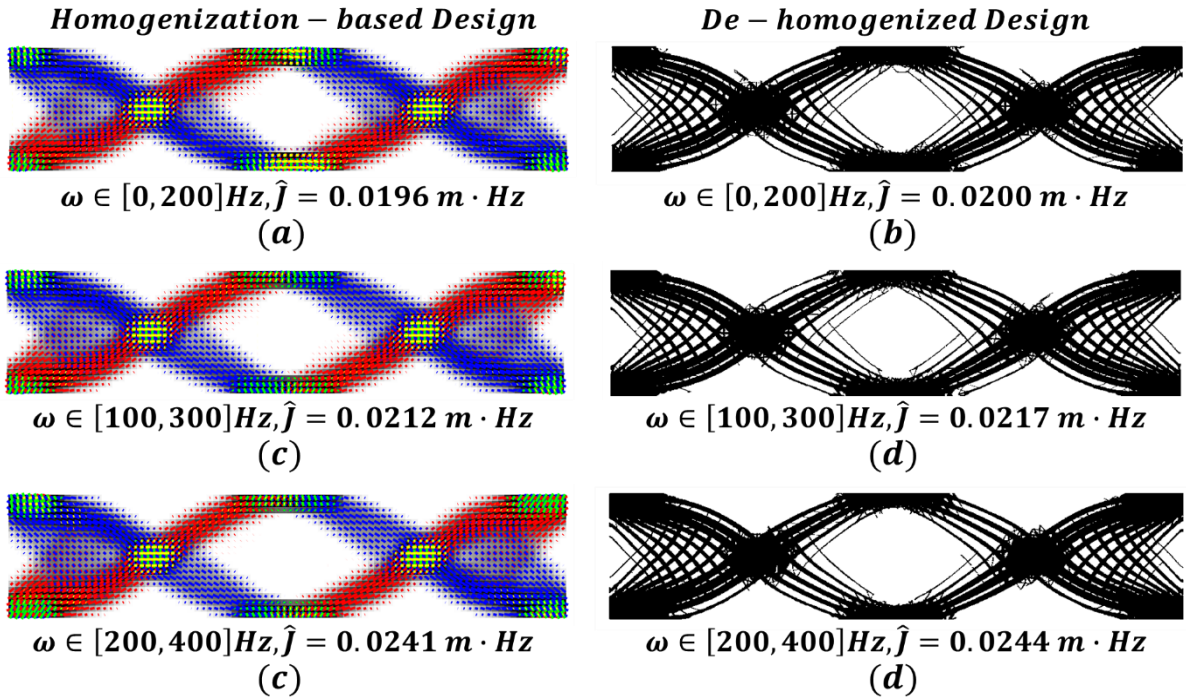

Fig. 15 Optimization results for the clamped beams given different frequency ranges of interest, (a), (c), (e) are homogenization-based optimization results on the coarse mesh for frequency ranges $[0,200]\ Hz, [100, 300]\ Hz$ and $[300, 400]\ Hz$, respectively; (b), (d), (f) are de-homogenized designs inferred from the optimization results in (a), (c), and (e), respectively.

From Fig. 15, it can be noticed that the proposed de-homogenized method can attain approximately the same frequency response performance as that of homogenization-based designs. Structures for different frequency ranges of interest have similar geometries. While designs for pinned beams in Fig. 11 form a loop-liked shape, these clamped designs are composed of one center loop and two mirrored half loops on both ends to better resist the clamped distortion. Despite the similarity in overall geometries, it should be noted that the bars will change their width and numbers when the frequency range of interest changes. Specifically, as the range of interest moves towards higher frequencies, the center loop will have its bars decrease in size and number, while bars in the two half loops go for the opposite. Interestingly, as shown in Fig. 16, we observe that this change in size and number of bars results in a sequential change of the order for the corresponding frequency response curves.

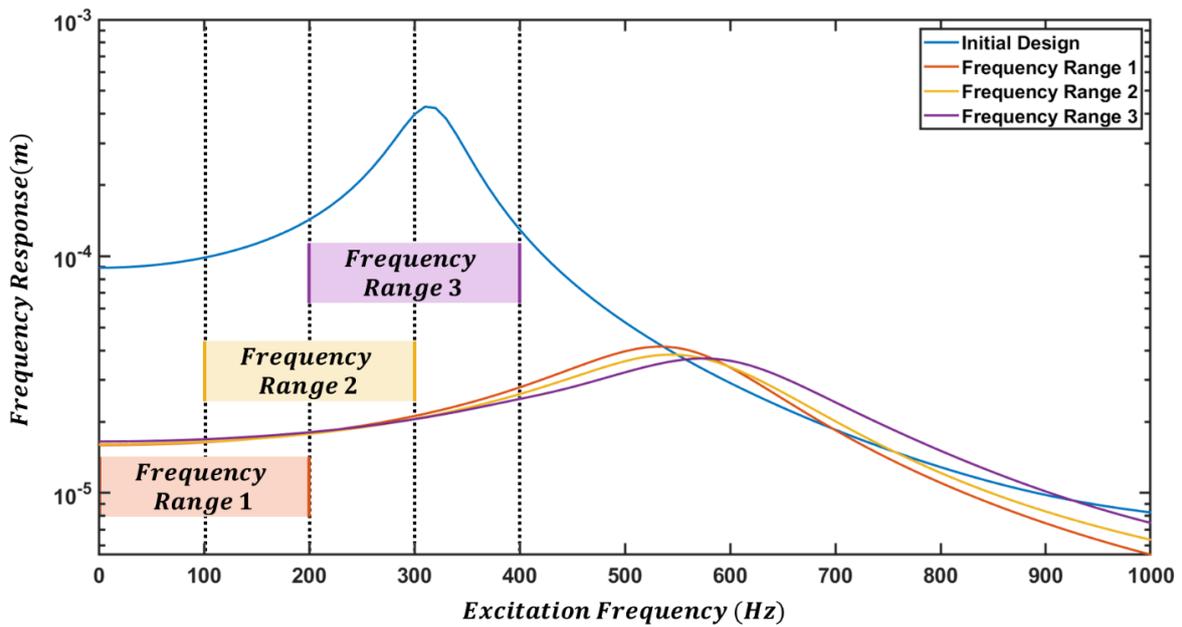

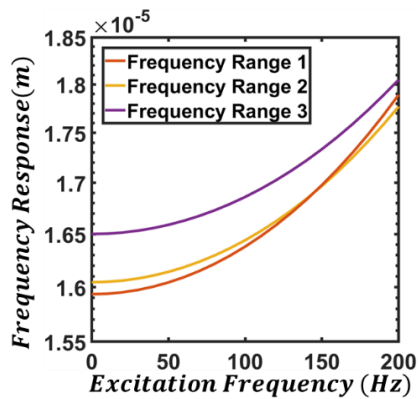 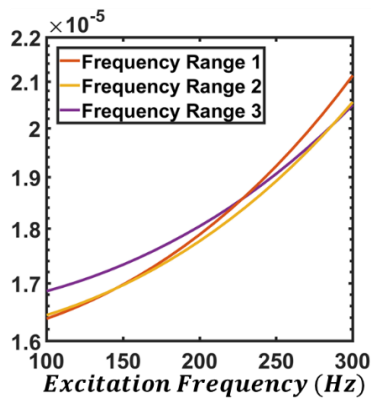 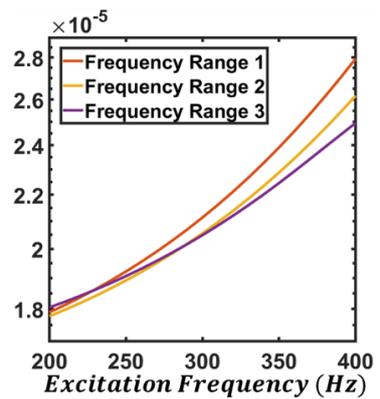

Fig. 16 Frequency response analysis of designs for different frequency ranges of interest. (a) frequency response within, (b)~(c) enlarged views for frequency responses of de-homogenized designs within different frequencies ranges.

Specifically, as shown in Figs. 16 (b) and (c), the structure (b) in Fig. 15 achieves the lowest level of response for most of the intervals in the first frequency range but exchanges its order with structure (d) in Fig. 15. This change is observed again, as shown in Fig. 16 (c) and (d), between structures (d) and (f) in Fig. 15 when the frequency goes from the second frequency range to the third one. As a result, each structure can only be optimal within its designed frequency range and become suboptimal for other frequency ranges. This also demonstrates the effectiveness of the proposed method in finding the optimal structures for a given frequency range.

In all of the previous designs, the regions of interest are placed symmetrically on the top and bottom of the beam. Combined with the symmetry of the loading condition, the corresponding homogenized designs also have symmetric geometries. Now, we design another multiscale structure to minimize the frequency response of the region on the bottom of the beam, but not for the one on the top, within a given frequency range of $[0,200]\ Hz$. As shown in Fig. 17(a) and (b), the de-homogenized structure has an asymmetric geometry with more bars in the lower half to reduce the distortion of the bottom, which can be indicated from the frequency response analysis shown in Fig. 17(c). The de-homogenized design can achieve dynamic performance close to the original homogenization-based design.

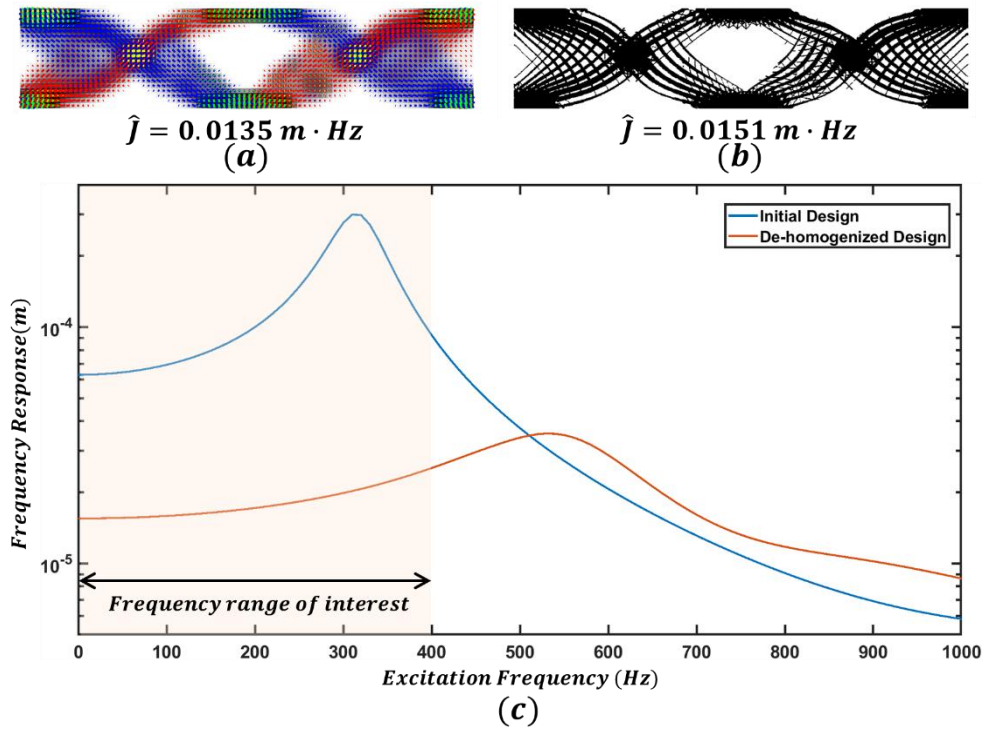

$\hat{J} = 0.0135\ m \cdot Hz$
(a)

$\hat{J} = 0.0151\ m \cdot Hz$
(b)

(c)

Fig. 17 Multiscale designs with asymmetric regions of interest. (a) homogenization-based optimization result, (b) the corresponding de-homogenized structure, and (c) frequency response analysis.

**4.3 Cantilever beam**

In this case study, we apply the proposed method to design a $2.4\ m \times 1.2\ m$ cantilever beam shown in Fig. 18. An excitation force with an amplitude of $2000N$ is imposed at the middle of the right end while its left end is fixed. The optimization objective is to minimize the frequency response of the loading point within a given excitation frequency range of $[0,200]\ Hz$. The beam is divided into an $80 \times 40$ mesh $\mathcal{H}_{ma}$ for optimization with 21 integration points in calculating the objective functions with equal subintervals. We obtained designs corresponding to three different volume fraction constraints, i.e., 40%, 50%, and 60%, as shown in Fig. 19.

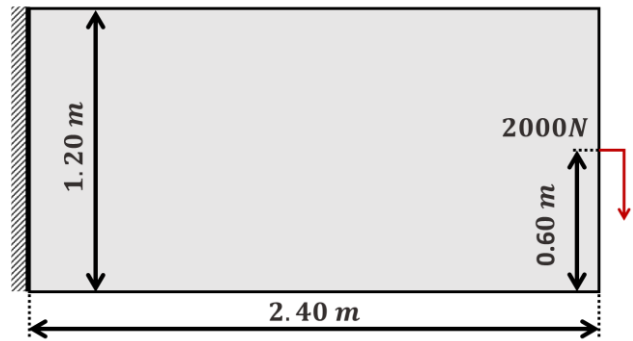

Fig. 18 Problem setting illustration of the third example.

From the results, it is noted that designs with different volume fraction constraints have similar multiscale structures. The orientation of the microstructures is consistent with the loading path of the macroscale structure. As the volume fraction of the solid increases, bars in the outer frame of the structure will become thicker. As a result, the value of the objective function decrease as the volume fraction increases. This can also be seen from the frequency response curves shown in Fig. 20, where the peak is moved further away and the response becomes lower in the frequency range of interest when the volume fraction increases. Moreover, the higher the volume fraction, the smaller the deviation of de-homogenized designs' performance from the corresponding homogenization-based design. This may be due to the inherent scaling deviation of 4-node quadrilateral mesh in dynamic simulation, and the thin bars discussed in the first case study. Nevertheless, all multiscale de-homogenized designs have very similar performance as the homogenization-based designs, demonstrating the effectiveness of the proposed method.

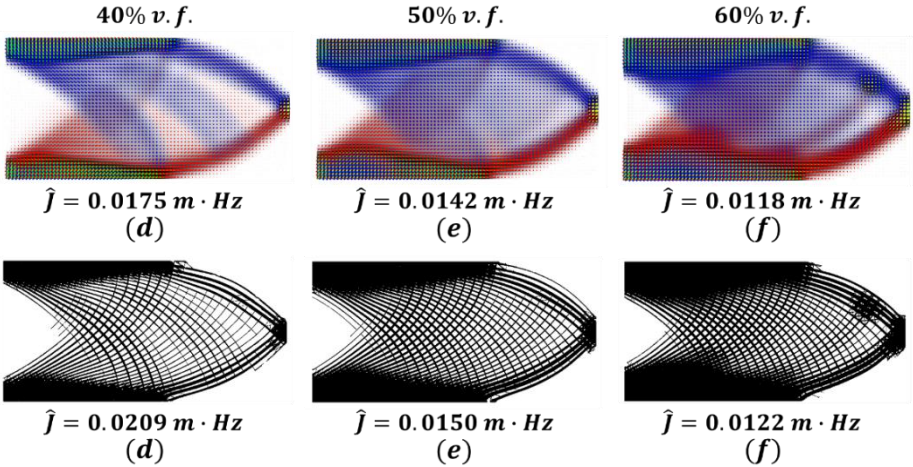

Fig. 19 Optimization results for the cantilever beams, (a)~(c) are homogenization-based optimization results under 40%, 50% and 60% volume fraction constraints, respectively; (d)~(f) are de-homogenized designs inferred from the design results in (a), (b), and (c), respectively.

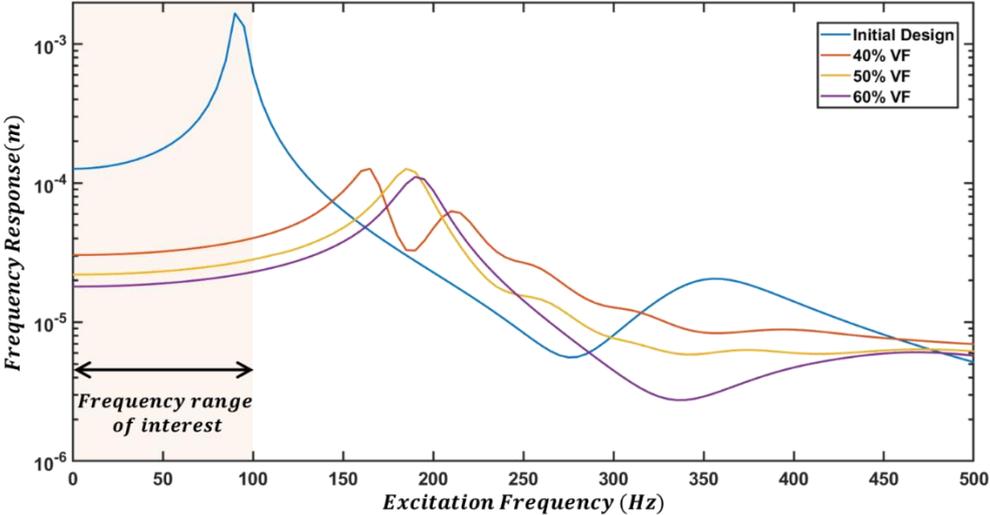

Fig. 20 Frequency response analysis for initial design and de-homogenized designs under different volume fraction constraints.

## 6. Conclusions

We have proposed a de-homogenization-based, data-driven topology optimization method to achieve efficient and effective multiscale designs for complex design cases. It allows the design of unit-cell orientation rather than simply assuming microstructures align with the principal strain direction. A sawtooth-function-based mapping is devised to expand the capability of de-homogenization to generate high-resolution multiscale structures with oriented microstructures instead of the usual square cell with rectangular holes. These techniques enable more flexible control of the microstructure properties to improve the applicability of de-homogenization in accommodating general design cases. Compared with the existing mapping methods for complex geometries, the proposed method can preserve the homogenized properties of unit-cells due to its conformality feature, and at the same time retain simplicity and efficiency.

With the proposed method, we have succeeded in applying the de-homogenization method to accommodate a flexible six-bar parameterized unit-cell representation for design applications beyond simple static compliance minimization, i.e., frequency response optimization. The de-homogenized structures obtained from the proposed multiscale method can achieve approximately the same dynamic performance as the homogenization-based designs, illustrating the effectiveness of the de-homogenization process in preserving the homogenized properties distribution. Through design cases, we have demonstrated that the proposed method can generate high-resolution multiscale structures with efficiency comparable to single-scale macroscale design but with much better dynamic performance. We observe that the multiscale structures have consistent main load-bearing directions to resist distortions under dynamic excitations. Moreover, the size of the unit-cells can be easily tuned by modifying the periodicity parameters of the sawtooth function. The proposed method can also adaptively change the structural geometries to optimize the frequency response of different regions and accommodate different excitation frequency ranges as well as volume fraction constraints.

The present study does not consider the singularities that might occur in the optimized orientation fields. However, the proposed sawtooth-function-based mapping could be easily integrated into existing de-homogenization methods that take singularities into consideration [31, 38]. Currently, the proposed mapping method can only ensure conformality for rectangular cells. Further studies are needed to enable conformal mapping for other close-packed tiling patterns, such as triangles, parallelograms and hexagons. Note that, even though we use six-bar unit-cells for demonstration in this study, the same approach can be readily applied to free-form rectangular cells with free-form microstructures. Although we only applied the proposed method for frequency response optimization in this study, it can be readily extended to a wide variety of static and dynamic design cases, such as those that consider thermal behaviors [56, 57], permeability [58], and fracture resistance [59,60]. We believe these applications would benefit more from the orientation design and the use of free-formed microstructures achieved by the proposed method.

**Acknowledgments**

We are grateful for support from the NSF CSSI program (Grant No. OAC 1835782) and Natural Science Foundation of Shanghai (21ZR1431500). Liwei Wang acknowledges support from the Zhiyuan Honors Program for Graduate Students of Shanghai Jiao Tong University for his predoctoral visiting study at Northwestern University. Yu-Chin Chan thanks the NSF Graduate Research Fellowship (Grant No. DGE-1842165).

**Appendix A. Details of the neural network structure**

In this work, a multi-layer neural network is used as the surrogate model with a four-dimensional input vector and a seven-dimensional output vector. It is composed of four hidden layers with the details shown in Table A1.

**Table A1.** Details of the neural network structure

| Layer name | Dim. of inputs | Dim. of outputs | Activation function |
|---|---|---|---|
| Input layer | 4 | 4 | \ |
| Hidden layer 1 | 4 | 8 | tanh |
| Hidden layer 2 | 8 | 16 | tanh |
| Hidden layer 3 | 16 | 32 | tanh |
| Hidden layer 4 | 32 | 7 | tanh |
| Output layer | 7 | 7 | linear |

**Appendix B. Theoretical proof**

**Proposition** *Sawtooth-function-based mapping $\psi(y)$ proposed in Equations (31) and (32) is an angle-preserving (conformal) mapping, given that P is a positive constant.*

*Proof*: The Jacobian matrix of the mapping can be obtained as

$$Jac_\psi(y) = \begin{bmatrix} \frac{\partial \psi_1}{\partial y_1} & \frac{\partial \psi_1}{\partial y_2} \\ \frac{\partial \psi_2}{\partial y_1} & \frac{\partial \psi_2}{\partial y_2} \end{bmatrix} = \begin{bmatrix} \frac{2}{P}\cos(\theta) & \frac{2}{P}\sin(\theta) \\ \frac{2}{P}\cos(\theta + \frac{\pi}{2}) & \frac{2}{P}\sin(\theta + \frac{\pi}{2}) \end{bmatrix}. \tag{A1}$$

It can be further simplified as

$$Jac_\psi(y) = \frac{2}{P} \cdot \begin{bmatrix} \cos(\theta) & \sin(\theta) \\ -\sin(\theta) & \cos(\theta) \end{bmatrix}. \tag{A2}$$

Substituting $\vartheta = -\theta$ into (A2), the Jacobian matrix can be formulated as

$$Jac_\psi(y) = \frac{2}{P} \cdot \begin{bmatrix} \cos(\vartheta) & -\sin(\vartheta) \\ \sin(\vartheta) & \cos(\vartheta) \end{bmatrix}, \tag{A3}$$

which can be viewed as a scaled multiple of a rotation matrix. Therefore, the mapping $\psi(y)$ is angle-preserved (conformal). ∎